\documentclass[smallextended]{svjour3}       
\smartqed  
\usepackage{graphicx,epsfig}

\begin{document}

\title{
On the quantum discord of general {\em X} states
}

\author{
M.A.Yurischev
}


\institute{
        M.~A.~Yurischev   \at
              Institute of Problems of Chemical Physics of the Russian Academy of Sciences,
Chernogolovka, 142432, Moscow Region, RUSSIA\\     
           \email{yur@itp.ac.ru} }

\date{Received:}

\maketitle

\begin{abstract}
Quantum discord $Q$ is a function of density matrix elements.
The domain of such a function in the case of two-qubit system with $X$
density matrix may consist of three subdomains at most:
two ones where the quantum discord is expressed in closed analytical
forms ($Q_{\pi/2}$ and $Q_0$) and an intermediate subdomain for which, to extract
the quantum discord $Q_\theta$, it is required to solve in general numerically
a one-dimensional minimization problem to find the optimal measurement angle
$\theta\in(0,\pi/2)$.
Hence the quantum discord is given by a piecewise-analytic-numerical formula
$Q=\min\{Q_{\pi/2},Q_\theta,Q_0\}$.
Equations for determining the boundaries between these subdomains are obtained.
The boundaries consist of bifurcation points.
The $Q_\theta$ subdomains are discovered in the generalized Horodecki states,
in the dynamical phase flip channel model,
in the anisotropic spin systems at thermal equilibrium,
in the heteronuclear dimers in an external magnetic field.
We found that transitions between $Q_\theta$ subdomain  and $Q_{\pi/2}$ and $Q_0$
ones occur suddenly but continuously and smoothly, i.e., nonanalyticity is hidden
and can be observed in higher derivatives of discord function.
\end{abstract}

\PACS{03.65.Ud \and 03.67. -a \and 75.10.Jm}

\keywords{$X$ density matrix \and quantum discord \and bifurcation points
 \and sudden transitions}

\section{Introduction}
\label{sec:Intro}
At present we have a situation where further miniaturization of electronics
will inevitably lead to molecular size components.
Designing such components requires application of the laws of quantum mechanics.
This is expected to lead to a technological breakthrough which will be achieved
through employing the holy of holies of the quantum theory --- so-called quantum
correlations.

Initially, the entanglement has been considered as a quantum correlation
\cite{AFOV08,HHHH09}.
Quantum entanglement is able to bind different parts of systems, even in the
case when there is no interaction between those parts (the Einstein-Podolsky-Rosen
effect).
Quantum entanglement exists only in nonseparable states of bi- and multipartite
systems. 
However, it appears in the last years that there are quantum
correlations more general and more fundamental than entanglement.
In particular, they can be present in certain separable
states, i.e., when the quantum entanglement is absent.
As a measure of total purely quantum correlations in bipartite systems,
the quantum discord is employed now \cite{CMS11,MBCPV12,AFY14}.
The basis for the discord conception is the idea of measurements
performed on a system and maximum amount of classical information
being extracted with their help.

Due to the fact that it is necessary to solve the optimization problem, the evaluation of
quantum correlations, especially discord, is extremely hard \cite{H14}.
If for the two-qubit systems the quantum entanglement of formation has been obtained for
the arbitrary density matrices \cite{W98}, the analytical formulas for the
quantum discord were proposed for $X$ states
\cite{Luo08,ARA10,FWBAC10,LWF11,DWZ11,VR12}.
In an $X$ matrix, nonzero entries may belong only to the main diagonal and
anti-diagonal \cite{YE07,R09}.
Notice that the sum and product of $X$ matrices is again the $X$ matrix
(i.e., a set of $X$ matrices is algebraically closed).

However, it was found later that
the formulas \cite{ARA10,FWBAC10,LWF11,DWZ11} are incorrect in general.
The reason is that the authors \cite{ARA10,FWBAC10,LWF11,DWZ11} believed (and it
was their error) that the optimal measurements are achieved only in the limiting
points, i.e., at the angles $\theta=0$ or $\pi/2$ (see below).
But on the explicit  examples \cite{LMXW11,CZYYO11,H13} of $X$ density matrices
it was proved that the optimal measurements can take place at the
intermediate angles in the interval $(0,\pi/2)$.
Unfortunately, these examples with density matrices are specific and do not clarify
the general situation.

In the present paper we show that the domain of intermediate optimal angles
can arise in the vicinity
of transition from the domain with optimal measurement angle $\theta=\pi/2$
to the domain with optimal angle $\theta=0$ (or inversely).
We derive the equations for the boundaries between these domains and
investigate their solutions for different models.
In particular, the boundaries can coincide or be absent
at all and then the quantum discord is given in the total domain of definition
by closed analytical formulas.

In the following sections, the general seven-parameters $X$ density matrix is reduced
to the five-parameter form by using local unitary transformations,
the existence of intermediate subdomains with the optimal anglers
$\theta\not=0, \pi/2$ is proved,
the equations for boundaries between different subdomains are derived
and then applied to various physical systems.
Finally, a brief conclusion is given.

\section{Real non-negative form for the $X$ density matrices.
Domain of definition for the matrix elements}
\label{sec:X-D}
In the most general case, the $X$ density matrix of two-qubit ($A$ and $B$ or 1 and 2)
system is given as
\begin{equation}
   \label{eq:rho-Xc}
   \rho_{AB}
	 =\left(
      \begin{array}{cccc}
      a&0&0&u_1+iu_2\\
      0&b&v_1+iv_2&0\\
      0&v_1-iv_2&c&0\\
      u_1-iu_2&0&0&d
      \end{array}
   \right), 
\end{equation}
where $a+b+c+d=1$.
This matrix contains seven real parameters which, due to the non-negativity
definition of any density operator, must satisfy the inequalities
\begin{equation}
   \label{eq:ineqs}
   a, b, c, d\ge0,\quad ad\ge u_1^2+u_2^2, \quad bc\ge v_1^2+v_2^2.
\end{equation}

The quantum entanglement and quantum discord are invariant under the local
unitary transformations of density matrices
\cite{AFOV08,HHHH09,CMS11,MBCPV12,AFY14}.
Owing to this property, one can with the help of transformation
\begin{equation}
   \label{eq:U}
   U=e^{-i\varphi_1\sigma_z/2}\otimes e^{-i\varphi_2\sigma_z/2}
\end{equation}
reduce the seven-parameters density matrix (\ref{eq:rho-Xc}) to the
real five-parameters $X$ form \cite{H13,CRC10,Y14,Y14a},
$U\rho_{AB}U^+\to\rho_{AB}$.
This provides with the angles
\begin{equation}
   \label{eq:phi12}
   \varphi_{1,2}=\frac{1}{2}(\arctan\frac{u_2}{u_1}\pm\arctan\frac{v_2}{v_1}).
\end{equation}
After this, the density matrix (\ref{eq:rho-Xc}) takes the form
\begin{equation}
   \label{eq:rho-Xr}
   \rho_{AB}
	 =\left(
      \begin{array}{cccc}
      a&0&0&u\\
      0&b&v&0\\
      0&v&c&0\\
      u&0&0&d
      \end{array}
   \right), 
\end{equation}
where 
\begin{equation}
   \label{eq:u}
   u=u_1\cos(\arctan\frac{u_2}{u_1}) + u_2\sin(\arctan\frac{u_2}{u_1}), 
\end{equation}
\begin{equation}
   \label{eq:v}
   v=v_1\cos(\arctan\frac{v_2}{v_1}) + v_2\sin(\arctan\frac{v_2}{v_1}). 
\end{equation}
Moreover, with the help of local rotations again around the $z$ axis, it is
not difficult to obtain also the non-negative off-diagonal elements of the
$X$ matrix (\ref{eq:rho-Xr}).
Indeed, the local unitary transformation
\begin{equation}
   \label{eq:U1}
   U_1=e^{-i\frac{\pi}{4}\sigma_z}\otimes e^{i\frac{\pi}{4}\sigma_z}
   =\left(
      \begin{array}{cccc}
      i&&&\\
      &1&&\\
      &&1&\\
      &&&-i
      \end{array}
   \right), 
\end{equation}
changes the sign of the off-diagonal matrix element $u$:
\begin{equation}
   \label{eq:rho-U1}
   U_1\rho_{AB} U_1^+
	 =\left(
      \begin{array}{rrrr}
      a&0&0&-u\\
      0&b&v&0\\
      0&v&c&0\\
      -u&0&0&d
      \end{array}
   \right). 
\end{equation}
Similarly, the local transformation
\begin{equation}
   \label{eq:U2}
   U_2=e^{i\frac{\pi}{4}\sigma_z}\otimes e^{-i\frac{\pi}{4}\sigma_z}
   =\left(
      \begin{array}{cccc}
      1&&&\\
      &i&&\\
      &&-i&\\
      &&&1
      \end{array}
   \right), 
\end{equation}
selectively acts on the sign of $v$:
\begin{equation}
   \label{eq:rho-U2}
   U_2\rho_{AB} U_2^+
	 =\left(
      \begin{array}{rrrr}
      a&0&0&u\\
      0&b&-v&0\\
      0&-v&c&0\\
      u&0&0&d
      \end{array}
   \right). 
\end{equation}
Thus, after transformation of $X$ matrix to the real form we may simply enclose
the off-diagonal elements in the modul symbols:
\begin{equation}
   \label{eq:rho-Xm}
   \rho_{AB}
	 =\left(
      \begin{array}{cccc}
      a&0&0&|u|\\
      0&b&|v|&0\\
      0&|v|&c&0\\
      |u|&0&0&d
      \end{array}
   \right). 
\end{equation}
This operation does not influence on the value of quantum correlations
in the system.

Thus, one can now consider the density matrices (\ref{eq:rho-Xr}) with restrictions
\begin{equation}
   \label{eq:ineqs1}
   a, b, c, d, u, v\ge0,\quad a+b+c+d=1, \quad ad\ge u^2, \quad bc\ge v^2.
\end{equation}
These relations define the domain ${\cal D}$ of $X$ density matrix in the space
of its entries.

We can rewrite the density matrix (\ref{eq:rho-Xr}) in the equivalent form
\begin{equation}
   \label{eq:rho-X1}
   \rho_{AB}
	 ={1\over4}\!\left(
      \begin{array}{cccc}
      1+s_1+s_2+c_3&.&.&c_1-c_2\\
      .&1+s_1-s_2-c_3&c_1+c_2&.\\
      .&c_1+c_2&1-s_1+s_2-c_3&.\\
      c_1-c_2&.&.&1-s_1-s_2+c_3
      \end{array}
   \right),
\end{equation}
where
\begin{eqnarray}
   \label{eq:s1-c3}
   &&s_1=a+b-c-d,\quad s_2=a-b+c-d,
   \nonumber\\
	 &&c_1=2(v+u),\quad c_2=2(v-u),\quad c_3=a-b-c+d.
\end{eqnarray}
Decomposition of this matrix on the Pauli matrices
$\sigma_\alpha$ ($\alpha=x,y,z$) leads to its Bloch form
\begin{equation}
   \label{eq:rhoXr-Bloch}
   \rho_{AB}=\frac{1}{4}(1
   + s_1\sigma_z\otimes1
   + s_21\otimes\sigma_z
   + c_1\sigma_x\otimes\sigma_x 
   + c_2\sigma_y\otimes\sigma_y 
   + c_3\sigma_z\otimes\sigma_z).
\end{equation}
The expansion coefficients are the unary and binary correlation functions
and therefore five parameters of density matrix are expressed through the
five different correlators,
\begin{eqnarray}
   \label{eq:corr-s1c3}
   &&s_1=\langle\sigma^1_z\rangle={\rm Tr}(\rho_{AB}\sigma_z\otimes1),\quad
   s_2=\langle\sigma^2_z\rangle={\rm Tr}(\rho_{AB}1\otimes\sigma_z),
   \nonumber\\
   &&c_1=\langle\sigma^1_x\sigma^2_x\rangle={\rm Tr}(\rho_{AB}\sigma_x\otimes\sigma_x),\quad
   c_2=\langle\sigma^1_y\sigma^2_y\rangle={\rm Tr}(\rho_{AB}\sigma_y\otimes\sigma_y),
   \\
   &&c_3=\langle\sigma^1_z\sigma^2_z\rangle={\rm Tr}(\rho_{AB}\sigma_z\otimes\sigma_z).
   \nonumber
\end{eqnarray}
It is clear that
\begin{equation}
   \label{eq:s1c3}
   -1\le s_1, s_2, c_1, c_2, c_3\le 1.
\end{equation}

The domain of definition, ${\cal D}$, in the space $(s_1, s_2, c_1, c_2, c_3)$
is formed, according to Eqs.~(\ref{eq:ineqs1}) and (\ref{eq:s1-c3}), by
conditions (see also \cite{CZYYO11,KHJP10})
\begin{equation}
   \label{eq:surfaces}
   (1 - c_3)^2\ge(s_1 - s_2)^2 + (c_1 + c_2)^2,\quad
   (1 + c_3)^2\ge(s_1 + s_2)^2 + (c_1 - c_2)^2. 
\end{equation}
The solid ${\cal D}$ is finite and lies in the five-dimensional
hypercube (\ref{eq:s1c3}).
Numerical calculations show that the volume of ${\cal D}$ is $8\%$
of the hypercube one.

The domain ${\cal D}$ is bounded by two quadratic hypersurfaces
\begin{equation}
   \label{eq:surf1}
   (s_1 - s_2)^2 + (c_1 + c_2)^2 - (c_3-1)^2 = 0
\end{equation}
and
\begin{equation}
   \label{eq:surf2}
   (s_1 + s_2)^2 + (c_1 - c_2)^2 - (c_3+1)^2 = 0. 
\end{equation}
Rotation by the angle $\pi/4$ around the $c_3$ axis transforms these
hyperquadrics to the forms
\begin{equation}
   \label{eq:surf1a}
   (s_2^\prime)^2 + (c_1^\prime)^2 - \frac{(c_3-1)^2}{2} = 0
\end{equation}
and
\begin{equation}
   \label{eq:surf2a}
   (s_1^\prime)^2 + (c_2^\prime)^2 - \frac{(c_3+1)^2}{2} = 0,
\end{equation}
where
\begin{equation}
   \label{eq:sc1}
   s_{1,2}^\prime = (\pm s_1+s_2)/\sqrt2,\quad
   c_{1,2}^\prime = (\pm c_1+c_2)/\sqrt2.
\end{equation}
Thus, the five-dimensional domain ${\cal D}$ results from an intersection of two conic
hypercylinders (\ref{eq:surf1a}) and (\ref{eq:surf2a}).

\section{Three alternatives for the quantum discord}
\label{sec:3Q}

As mentioned above, the measurement operations lie in the ground of discord
notion.
Following the founders of discord conception \cite{OZ01,Z03} and their adherents
\cite{Luo08,ARA10,FWBAC10,LWF11,DWZ11} we will consider here the one-dimensional
projective measurements (i.~e., the von Neumann measurements).
Such measurements can be reduced to projections which are characterized
by the polar ($\theta$) and azimuthal ($\phi$) angles relative to the $z$ axis.
It is important that in the case of real $X$ density matrix with an additional
condition $uv\ge0$ the optimal measurements are achieved by $\cos2\phi=1$
\cite{H13,CRC10}.
Since the sign of off-diagonal elements are changed by the local unitary transformations,
we can always satisfy the above condition.

In the general case, quantum discord depends on subsystem ($A$ or $B$) where the measurements
are performed.
For definiteness and without loss of generality, let the measured subsystem be $B$.
Then the quantum discord is given as \cite{CMS11,MBCPV12,AFY14}
\begin{equation}
   \label{eq:Q}
   Q=S(\rho_B)-S(\rho_{AB})+\min_\theta S_{cond}(\theta),
\end{equation}
where $\rho_B={\rm Tr}_A\rho_{AB}$ is the reduced density matrix and
$S(\rho)=-{\rm Tr}\rho\ln\rho$ is the von Neumann entropy
for the corresponding state $\rho$.
(Here the entropy is in nats; to transform it, e.g., in bits, one should divide
it by $\ln2$.)
Simple calculations with (\ref{eq:rho-Xr}) lead to
\begin{equation}
   \label{eq:SB}
   S(\rho_B)=-(a+c)\ln(a+c)-(b+d)\ln(b+d),
\end{equation}
$S(\rho_{AB})=S$, where
\begin{eqnarray}
   \label{eq:S}
	 &&S=-{a+d+\sqrt{(a-d)^2+4u^2}\over2}\ln{a+d+\sqrt{(a-d)^2+4u^2}\over2}
   \nonumber\\
	 &&-{a+d-\sqrt{(a-d)^2+4u^2}\over2}\ln{a+d-\sqrt{(a-d)^2+4u^2}\over2}
   \nonumber\\
	 &&-{b+c+\sqrt{(b-c)^2+4v^2}\over2}\ln{b+c+\sqrt{(b-c)^2+4v^2}\over2}
   \nonumber\\
	 &&-{b+c-\sqrt{(b-c)^2+4v^2}\over2}\ln{b+c-\sqrt{(b-c)^2+4v^2}\over2}.
\end{eqnarray}
The quantum conditional entropy of subsystem $A$ is given as \cite{H13} 
\begin{equation}
   \label{eq:Scond}
   S_{cond}(\theta)=\Lambda_1\ln\Lambda_1+\Lambda_2\ln\Lambda_2
	 -\sum_{i=1}^4\lambda_i\ln\lambda_i, 
\end{equation}
where
\begin{equation}
   \label{eq:Lam12}
   \Lambda_{1,2}={1\over2}[1\pm (a-b+c-d)\cos\theta],
\end{equation}
\begin{eqnarray}
   \label{eq:lam12}
   \lambda_{1,2}&=&{1\over4}\lbrack\!\lbrack1+(a-b+c-d)\cos\theta
   \nonumber\\
	 &\pm&\{[a+b-c-d+(a-b-c+d)\cos\theta]^2+4w^2\sin^2\theta\}^{1/2}\rbrack\!\rbrack,
\end{eqnarray}
\begin{eqnarray}
   \label{eq:lam34}
   \lambda_{3,4}&=&{1\over4}\lbrack\!\lbrack1-(a-b+c-d)\cos\theta
   \nonumber\\
	 &\pm&\{[a+b-c-d-(a-b-c+d)\cos\theta]^2+4w^2\sin^2\theta\}^{1/2}\rbrack\!\rbrack,
\end{eqnarray}
\begin{equation}
   \label{eq:w}
   w=|u|+|v|.
\end{equation}
Thus, $S_{cond}$ depends in fact on four parameters because $u$ and $v$ enter
via the combination (\ref{eq:w}).
The conditional entropy $S_{cond}(\theta)$ is a differentiable function
of its argument $\theta$. 

\begin{figure}[t]
\begin{center}
\epsfig{file=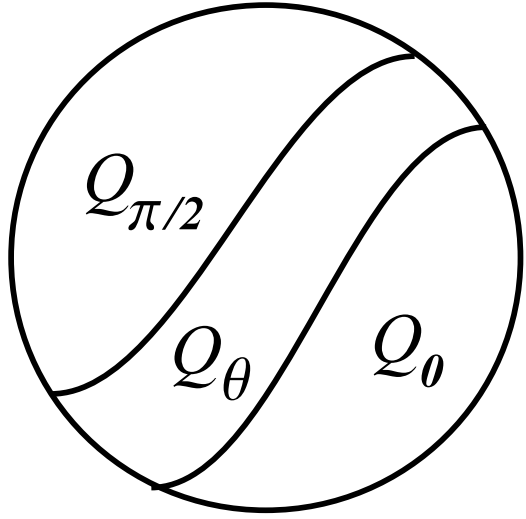,width=6.8cm}
\caption{A fragment of phase diagram with three possible subdomains
for the $X$-state quantum discord}
\label{fig:1}
\end{center}
\end{figure}

Expressions (\ref{eq:SB})-(\ref{eq:lam34}) allow to define the measurement-dependent
discord as \cite{MBCPV12}
\begin{equation}
   \label{eq:Q-theta}
   Q(\theta)=S(\rho_B)-S(\rho_{AB})+S_{cond}(\theta),
\end{equation}
where $\theta\in[0,\pi/2]$.
It is obvious that the absolute minimum of this discord can be either on the bounds
($\theta=0,\pi/2$) or at the intermediate point $\theta\in(0,\pi/2)$.
As a result, there is a choice from three possibilities
for the quantum discord
\begin{equation}
   \label{eq:Q3}
   Q=\min\{Q_0, Q_\theta, Q_{\pi\over2}\}.
\end{equation}
This equation generalizes the one proposed earlier for the quantum discord
\cite{Luo08,ARA10,FWBAC10,LWF11,DWZ11}
\begin{equation}
   \label{eq:Q2}
   \tilde Q=\min\{Q_0, Q_{\pi\over2}\},
\end{equation}
i.e., it was assumed that the optimal observable can be either
$\sigma_z$ or $\sigma_x$.
In Fig.~\ref{fig:1} we schematically illustrate the parameter domain of a system
with three possible subdomains for the discord.

From Eqs.~(\ref{eq:SB})-(\ref{eq:Q-theta}), we have for the
discord branch $Q_0\equiv Q(0)$:
\begin{equation}
   \label{eq:Q0}
   Q_0=-S-a\ln a-b\ln b-c\ln c -d\ln d.
\end{equation}
For $\theta=\pi/2$ we obtain
\begin{eqnarray}
   \label{eq:Q1}
   &&Q_{\pi\over2}=-S-(a+c)\ln(a+c)-(b+d)\ln(b+d)
   \nonumber\\
	 &&-{1+\sqrt{(a+b-c-d)^2+4w^2}\over2}\ln{1+\sqrt{(a+b-c-d)^2+4w^2}\over2}
   \nonumber\\
	 &&-{1-\sqrt{(a+b-c-d)^2+4w^2}\over2}\ln{1-\sqrt{(a+b-c-d)^2+4w^2}\over2}.
\end{eqnarray}
Thus, the branches $Q_0$ and $Q_{\pi/2}$ are expressed analytically,
and the branch $Q_\theta=\min\nolimits_{\theta\in(0,\pi/2)} Q(\theta)$,
if the intermediate minimum exists, should be found
from the numerical solution of one-dimensional minimization problem or from
the transcendental equation
\begin{equation}
   \label{eq:theta}
   S^\prime_{cond}(\theta)=0.
\end{equation}
The first derivative of conditional entropy with respect to $\theta$ is equal to
\begin{equation}
   \label{eq:ScondI}
   S^\prime_{cond}(\theta)=\Lambda^\prime_1(1+\ln\Lambda_1)
	 +\Lambda^\prime_2(1+\ln\Lambda_2)
	 -\sum_{i=1}^4\lambda^\prime_i(1+\ln\lambda_i)
\end{equation}
with
\begin{equation}
   \label{eq:Lam12I}
   \Lambda^\prime_{1,2}=\mp{1\over2}(a-b+c-d)\sin\theta,
\end{equation}
\begin{eqnarray}
   \label{eq:lam12I}
   &&\lambda^\prime_{1,2}={1\over4}\bigg[-(a-b+c-d)\sin\theta
   \nonumber\\
	 &&\pm\frac{[a+b-c-d+(a-b-c+d)\cos\theta][-(a-b-c+d)\sin\theta]+2w^2\sin2\theta}
	 {\sqrt{[a+b-c-d+(a-b-c+d)\cos\theta]^2+4w^2\sin^2\theta}}\bigg],
   \nonumber\\
\end{eqnarray}
\begin{eqnarray}
   \label{eq:lam34I}
   &&\lambda^\prime_{3,4}={1\over4}\bigg[(a-b+c-d)\sin\theta
   \nonumber\\
	 &&\pm\frac{[a+b-c-d-(a-b-c+d)\cos\theta](a-b-c+d)\sin\theta+2w^2\sin2\theta}
	 {\sqrt{[a+b-c-d-(a-b-c+d)\cos\theta]^2+4w^2\sin^2\theta}}\bigg].
   \nonumber\\
\end{eqnarray}
  
All three possible variants for the quantum discord ($Q_0$, $Q_{\pi/2}$, and $Q_\theta$)
can really exist in physical systems.
In the case when $a=b$ and $b=c$ (or $s_1=s_2=0$)
the conditional entropy minimum
is always achieved at one of two bound points \cite{Luo08}.
However, this is wrong for the more general $X$ states; global minimum can take place
at inner points of the interval $(0,\pi/2)$.

\begin{figure}[t]
\begin{center}
\epsfig{file=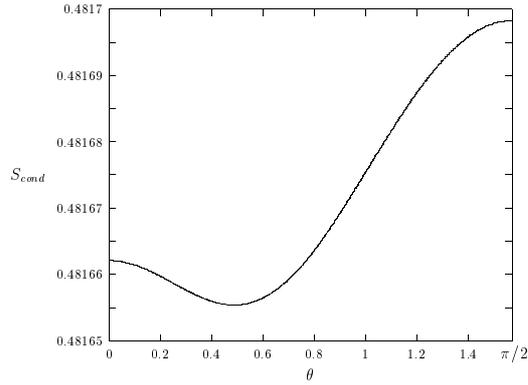,width=6.8cm}
\caption{Quantum conditional entropy $S_{cond}$ as a function of measured angle
$\theta$ for the state (\ref{eq:rho-LMXW})}
\label{fig:2}
\end{center}
\end{figure}

Indeed, following the authors \cite{LMXW11}, let us consider the state
\begin{equation}
   \label{eq:rho-LMXW}
   \rho
	 =\left(
      \begin{array}{cccc}
      0.0783&0&0&0\\
      0&0.125&0.100&0\\
      0&0.100&0.125&0\\
      0&0&0&0.6717
      \end{array}
   \right). 
\end{equation}
Using Eqs.~(\ref{eq:Scond})-(\ref{eq:w}) we have computed the function
$S_{cond}(\theta)$ for this state.
Its behavior is shown in Fig.~\ref{fig:2}.
From the figure we conclude that the conditional entropy minimum is situated in the
intermediate region, namely, at the angle $\theta=0.4883\approx28^\circ$.
Two other similar numerical examples of quantum states are given in Ref.~\cite{H13}.

These examples clearly show that the optimal measurement angles can really be
in the intermediate region $(0,\pi/2)$, i.e., the optimal observables for quantum
discord can be not only the $\sigma_x$ or $\sigma_z$, but also their superposition.

For the real $X$ state with constraint $|u+v|\ge|u-v|$
(i.e., $uv\ge0$ or ${\rm sign}u={\rm sign}v$),
the authors \cite{CZYYO11} have proved a theorem which guarantees that the optimal
observable is $\sigma_z$ if
\begin{equation}
   \label{eq:sz}
   (|u|+|v|)^2\le(a-b)(d-c)
\end{equation}
and $\sigma_x$ if
\begin{equation}
   \label{eq:sx}
   |u|+|v|\ge|\sqrt{ad}-\sqrt{bc}|.
\end{equation}
The theorem states nothing for the region lying between these bounds.
But in the case
\begin{equation}
   \label{eq:acbd}
    ac=bd
\end{equation}
the inequalities (\ref{eq:sz}) and (\ref{eq:sx}) lead to absence of the intermediate
region \cite{PKF13}.

\section{Equations for the boundaries}
\label{sec:bound}
Let us start with a heuristic example.
Consider a two-parameter family of $X$ states
\cite{CZYYO11,GGZ11}
\begin{equation}
   \label{eq:rho-me}
   \rho=\left(
      \begin{array}{cccc}
      \epsilon/2&0&0&\epsilon/2\\
      0&(1-\epsilon)m&0&0\\
      0&0&(1-\epsilon)(1-m)&0\\
      \epsilon/2&0&0&\epsilon/2
      \end{array}
   \right)
\end{equation}
or
\begin{eqnarray}
   \label{eq:rho-me-Bloch1}
   \rho&=&\frac{1}{4}[1
   + (1-\epsilon)(2m-1)(\sigma_1^z - \sigma_2^z)
   + \epsilon(\sigma_1^x\sigma_2^x - \sigma_1^y\sigma_2^y) 
   + (2\epsilon-1)\sigma_1^z\sigma_2^z]
	 \nonumber\\
	 &=&\epsilon|\Phi^+\rangle\langle\Phi^+|
	 +(1-\epsilon)m|01\rangle\langle01|+(1-\epsilon)(1-m)|10\rangle\langle10|,
\end{eqnarray}
where $|\Phi^+\rangle=(|00\rangle+|11\rangle)/\sqrt2$.
The given density matrix $\rho$ represents the generalized Horodecki states
\cite{KHJP10}.

Simple calculation yields $Q_0=\epsilon$ (in bits).
Sufficient conditions (\ref{eq:sz}) and (\ref{eq:sx}) for the $Q_0$
and $Q_{\pi/2}$ subdomains give \cite{CZYYO11}
\begin{equation}
   \label{eq:e0}
	 \epsilon\le\frac{2m(1-m)}{1+2m(1-m)}
\end{equation}
and
\begin{equation}
   \label{eq:e1}
	 \epsilon\ge\frac{\sqrt{m(1-m)}}{1+\sqrt{m(1-m)}},
\end{equation}
respectively.
But in the region
\begin{equation}
   \label{eq:e1e0}
	 \frac{2m(1-m)}{1+2m(1-m)}<\epsilon<\frac{\sqrt{m(1-m)}}{1+\sqrt{m(1-m)}}
\end{equation}
the above theorem does not say anything.

Let us now find the lines on the plane $(m, \epsilon)$ which are defined by
the condition
\begin{equation}
   \label{eq:Q0Q1}
   Q_0(m,\epsilon)=Q_{\pi\over2}(m,\epsilon).
\end{equation}
Then we will study the changes of curves $S_{cond}(\theta)$ in the
neighborhood to those lines.

\begin{figure}[t]
\begin{center}
\epsfig{file=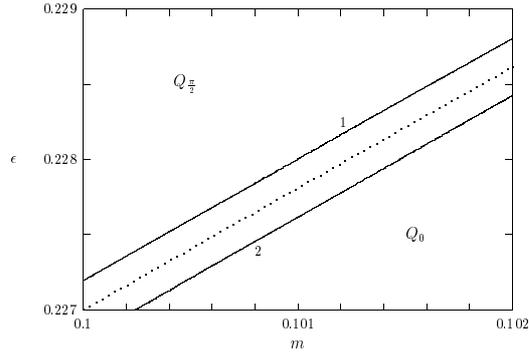,width=6.8cm}
\caption{Subdomains $Q_{\pi/2}$, $Q_0$, and (between them) $Q_\theta$ for the state
(\ref{eq:rho-me}).
Dotted line corresponds to the condition $Q_{\pi/2}=Q_0$.
Solid lines 1 and 2 are the $\pi/2$- and 0-boundaries, respectively}
\label{fig:3}
\end{center}
\end{figure}

Using Eqs.~(\ref{eq:S}), (\ref{eq:Q0}), and (\ref{eq:Q1})
we have numerically solved the transcendental equation (\ref{eq:Q0Q1}).
The solution is only one.
The results are plotted in Fig.~\ref{fig:3} by dotted line.

\begin{figure}[t]
\begin{center}
\epsfig{file=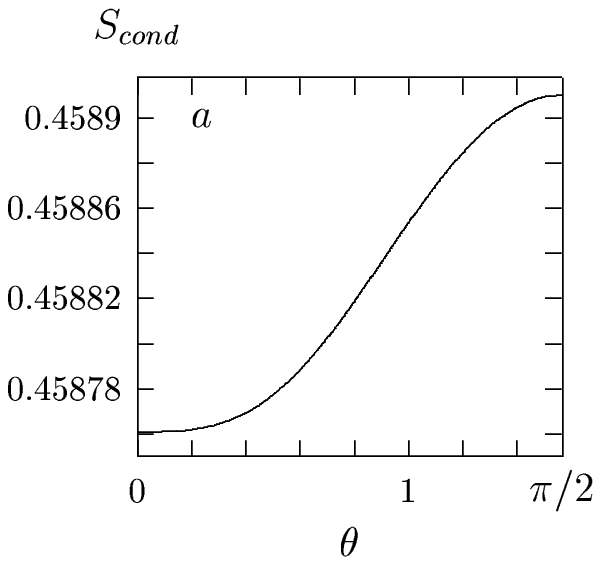,width=2.3cm}
\epsfig{file=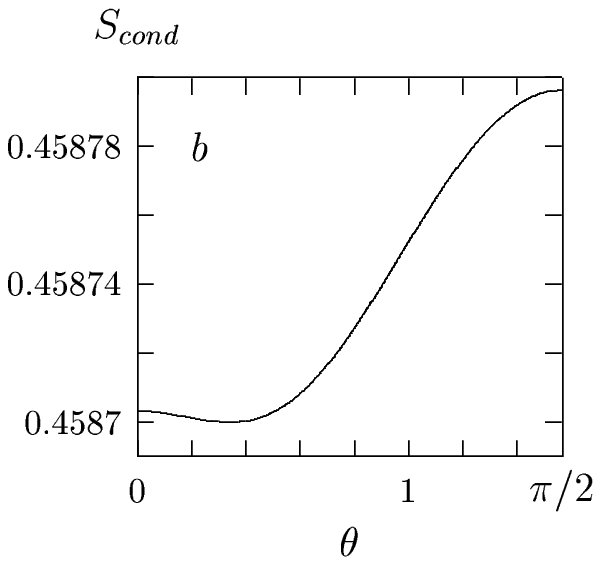,width=2.3cm}
\epsfig{file=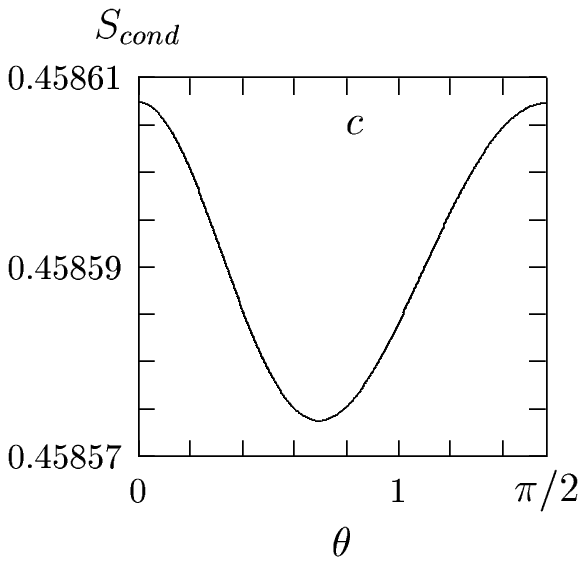,width=2.3cm}
\epsfig{file=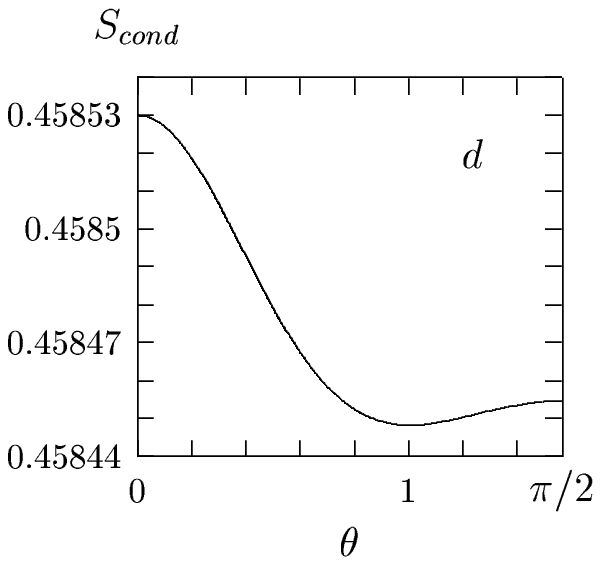,width=2.3cm}
\epsfig{file=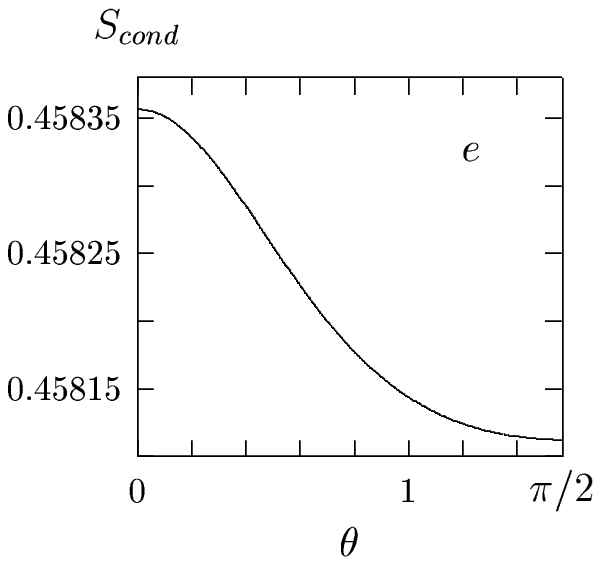,width=2.3cm}
\caption{
Appearance and disappearance of an intermediate minimum on the conditional entropy curve
by transition from $Q_0$ to $S_{\pi/2}$ subdomain.
Here, $S_{cond}(\theta)$ corresponds to the state (\ref{eq:rho-me}) at the fixed value
of $\epsilon=0.228$
and $m=0.1015~(a)$, $0.1014~(b)$, $0.101\,234~(c)$, $0.1011~(d)$,
and $0.1008~(e)$}
\label{fig:4}
\end{center}
\end{figure}

Consider in detail a particular case.
Let the $\epsilon$ is held fixed and equal, for example, to $\epsilon=0.228$
(see Fig.~\ref{fig:3}).
Then the equality $Q_0=Q_{\pi/2}$ is satisfied at the crossing point
$m_\times=0.101\,234$.
Study now the behavior of $S_{cond}(\theta)$ when the parameter $m$ varies.
Inequalities~(\ref{eq:e0}) and (\ref{eq:e1}) guarantee that when $m<0.096\,545$
the discord equals $Q=Q_{\pi/2}$ and $Q=Q_0$ when $m>0.180\,107$.
If $m=0.1015$, the minimum of $S_{cond}(\theta)$ is at $\theta=0$
[see Fig.~\ref{fig:4}~$(a)$].
Moreover, the angle $\theta=0$ is optimal for all larger values of $m$.
When the $m$ decreases the minimum on the curve
$S_{cond}(\theta)$ inside the interval between 0 and $\pi/2$ appears.
The minimum is clearly seen when $m=0.1014$ [Fig.~\ref{fig:4}~$(b)$].
Near the point $m=0.101\,234$ the minimum achieves large depth.
By further decreasing $m$, the minimum moves to the
bound $\theta=\pi/2$ and then it disappears at all.
Optimal measurements undergo to the angle $\theta=\pi/2$.

We argue now that both lower and upper boundaries of the interval where the
optimal angles lie between 0 and $\pi/2$ are exact,
i.e., the intermediate minimum of $S_{cond}(\theta)$ suddenly appears and suddenly
disappears.
Above all, we note that the first derivative of function $S_{cond}(\theta)$
at $\theta=0$ and $\pi/2$ equals zero in general case:
$S^{\prime}_{cond}(0)\equiv S^{\prime}_{cond}(\pi/2)\equiv0$.
This is easy to check by direct calculations using
Eqs.~(\ref{eq:ScondI})-(\ref{eq:lam34I}).
Let us turn now again to the Fig.~\ref{fig:4}.
By fixed value of parameter $\epsilon$ and for each value of $m$
one can say at any moment the inside minimum exists or it is absent.
For instance, when $m=0.1015$ ($\epsilon=0.228$) the function $S_{cond}(\theta)$ is
concave at the point $\theta=0$ and therefore its second derivative
$S^{\prime\prime}_{cond}(\pi/2)<0$.
But when $m=0.1014$ the conditional entropy has a local maximum at the same
bound point $\theta=0$ and therefore $S^{\prime\prime}_{cond}(\pi/2)>0$.
Hence, the bifurcation point (in the sense that two extrema arise from one) \cite{A04}
is determined by the
condition
\begin{equation}
   \label{eq:SII1}
   S^{\prime\prime}_{cond}(0)=0.
\end{equation}
Similarly we have for the other bound point $\theta=\pi/2$,
\begin{equation}
   \label{eq:SII0}
   S^{\prime\prime}_{cond}(\pi/2)=0.
\end{equation}
Using Eqs.~(\ref{eq:Scond})-(\ref{eq:lam34}) we get the second derivatives at
limiting points:
\begin{eqnarray}
   \label{eq:d2S0}
   &&S^{\prime\prime}_{cond}(0)={1\over4}(a-b+c-d)\biggl(2\ln\frac{b+d}{a+c}+\ln{ac\over bd}\biggr)
   \nonumber\\
	 &&+{1\over4}(a-b-c+d)\ln{ad\over bc}-{1\over2}w^2\biggl({1\over a-c}\ln{a\over c}
	 +{1\over b-d}\ln{b\over d}\biggr)
\end{eqnarray}
and
\begin{eqnarray}
   \label{eq:d2S1}
   &&S^{\prime\prime}_{cond}(\pi/2)
	 =\frac{8w^2}{r^3}[(a-c)(b-d)+w^2]\ln\frac{1+r}{1-r}+(a-b+c-d)^2
   \nonumber\\
	 &&-{1\over2(1+r)}\bigl[a-b+c-d+{1\over r}(a+b-c-d)(a-b-c+d)\bigl]^2
   \nonumber\\
	 &&-{1\over2(1-r)}\bigl[a-b+c-d-{1\over r}(a+b-c-d)(a-b-c+d)\bigl]^2 ,
\end{eqnarray}
where
\begin{equation}
   \label{eq:r}
   r=[(a+b-c-d)^2+4w^2]^{1/2}
\end{equation}
and $w$ is given by Eq.~(\ref{eq:w}).
The relations (\ref{eq:SII1})-(\ref{eq:r}) are the boundary equations for the
crossover subdomain $Q_\theta$ \cite{Y14,Y14a}.
Thus, the boundaries consist of bifurcation points.

If the solutions of Eqs.~(\ref{eq:SII1}) and (\ref{eq:SII0}) are the same,
the intermediate subdomain $Q_\theta$ is absent and the quantum discord is given by
analytical expressions.
On the other hand, instead of rough conditions (\ref{eq:sz}) and (\ref{eq:sx}),
the inequalities
  $S^{\prime\prime}_{cond}(0)\le0$
and
	$ S^{\prime\prime}_{cond}(\pi/2)\le0$
define now the complete subdomains $Q_0$ and $Q_{\pi/2}$, respectively.

Numerical solution of Eqs.~(\ref{eq:SII1})-(\ref{eq:r}) for the state (\ref{eq:rho-me})
shows that the boundaries are the lines going approximately parallel to the
dotted lines (see the lines 1 and 2 in Fig.~\ref{fig:3}).
As a result, the subdomain appears within which the optimal angles should
be found numerically.
Out of this subdomain we have analytical expressions for the quantum discord.
By $\epsilon=0.228$, the value for $m$ of $\pi/2$-boundary equals $m_{\pi/2}=0.100\,997$
and for the 0-boundary it is $m_0=0.101\,474$.
The middle of this interval equals 0.101\,236 which is near the point
$m_\times=0.101\,234$.

\begin{figure}[t]
\begin{center}
\epsfig{file=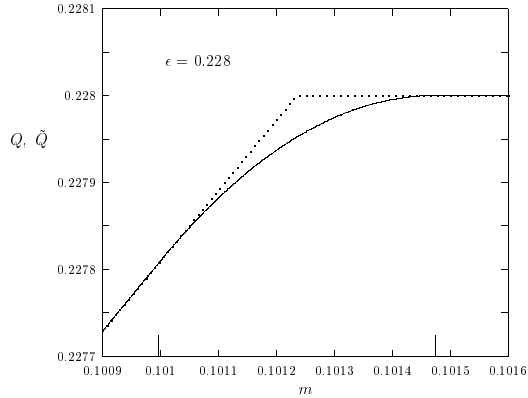,width=6.8cm}
\caption{Dependencies of the false discord $\tilde Q=\min\{Q_{\pi/2},Q_0\}$
(dotted line) and
the corrected quantum discord $Q=\min\{Q_{\pi/2},Q_\theta,Q_0\}$ (solid line).
for the state (\ref{eq:rho-me}) with parameter $\epsilon=0.228$
Longer bars mark the exact boundaries $m_{\pi/2}=0.100\,997$ and $m_0=0.101\,474$.
Subdomains $m\le m_{\pi/2}$, $m_{\pi/2}<m<m_0$, and $m\ge m_0$ correspond
to the discord branches $Q_{\pi/2}$, $Q_\theta$, and $Q_0$ respectively}
\label{fig:5}
\end{center}
\end{figure}

Consider the discord behavior by a transition from the subdomain
$Q_{\pi/2}$ to $Q_0$ one (Fig.~\ref{fig:5}).
One can see that down to crossing point $m_\times=0.101\,234$ the
discord ${\tilde Q}$, according to Refs.~\cite{ARA10,FWBAC10,LWF11}, equals
$Q_{\pi/2}$ and above the point $m_\times$ it equals $Q_0$
(see Fig.~\ref{fig:5}).
If this were valid, the discord ${\tilde Q}=\min\{Q_{\pi/2},Q_0\}$
would not be differentiable at the intersection point $m_\times$.
However, in fact, the true discord $Q=\min\{Q_{\pi/2},Q_\theta,Q_0\}$ is smooth.
This follows from the numerical solution of the task in the intermediate domain.
The results are shown again in Fig.~\ref{fig:5} by solid line.
It is clearly seen that smoothness occurs.
We may say that, instead a fracture at $m_\times$, two hidden transitions
occur at the $\pi/2$- and 0-boundaries.

Notice that the conditions (\ref{eq:e0}) and (\ref{eq:e1}) are
rough too and lead to the bounds which lie far beyond the region
of Fig.~\ref{fig:3}.

\section{Bell-diagonal states}
\label{subsec:Bd}
\begin{figure}[t]
\begin{center}
\epsfig{file=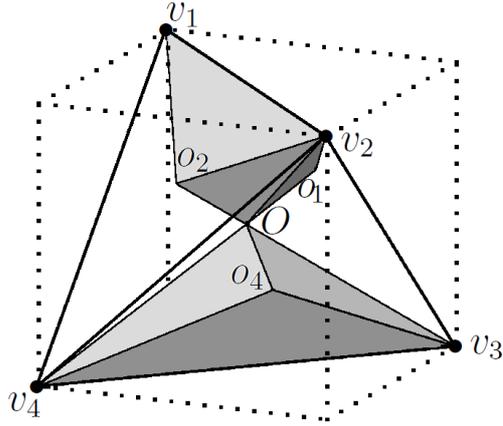,width=6.8cm}
\caption{
Tetrahedron with vertices $v_1$, $v_2$, $v_3$ and $v_4$ is the domain
for the total Bell-diagonal states.
Two regions $(O,v_1,v_2,o_1,o_2)$ and $(O,v_3,v_4,o_3,o_4)$ correspond
to the physical states with $Q_0$ discord
}
\label{fig:6}
\end{center}
\end{figure}

The case $a=d$ and $b=c$ or $s_1=s_2=0$ (i.e., when both local Bloch vectors are zero)
corresponds to the Bell-diagonal states.
Domain of definition for the physical states, ${\cal D}$, lies now in the
three-dimensional cube defined by $c_1, c_2, c_3\in[-1,1]$.
Two second-order hypersurfaces  (\ref{eq:surf1}) and  (\ref{eq:surf2}) are
transformed to the two first-order surfaces
\begin{equation}
   \label{eq:2plains1}
   \pm|c_1+c_2|+c_3-1=0
\end{equation}
and
\begin{equation}
   \label{eq:2plains2}
   \pm|c_1-c_2|+c_3+1=0 .
\end{equation}
The former consists of two semi-planes
with a $\wedge$-shaped cross section and the latter is similar to it but has
a $\vee$-shaped cross section.
The angle between semi-planes equals $\arccos(1/3)\approx78^\circ$.
These semi-plane surfaces put bounds to the domain ${\cal D}$ that is reduced,
as shown in Fig.~\ref{fig:6}, to a tetrahedron
with vertices \cite{HH96}
\begin{equation}
   \label{eq:4vertex}
   v_1=(-1,1,1),\quad v_2=(1,-1,1),\quad v_3=(1,1,-1),\quad v_4=(-1,-1,-1);
\end{equation}
these vertices lie in octants II $(-,+,+)$, IV $(+,-,+)$, V $(+,+,-)$,
and VII $(-,-,-)$, respectively.
The centers of tetrahedron facets are
\begin{eqnarray}
   \label{eq:4centers}
   &&o_1=(1/3,1/3,1/3),\quad o_2=(-1/3,-1/3,1/3),
   \nonumber\\
	 &&o_3=(-1/3,1/3,-1/3),\quad o_4=(1/3,-1/3,-1/3).
\end{eqnarray}
Tetrahedron volume equals a third (i.e., about $33.3\%$) of the cube one.
Notice that the tetrahedron vertices are the states with maximal value of discord
(which equals one in bit units).

It is known \cite{FACCA10} that the states with zero discord are negligible
in the whole Hilbert space.
In particular, it has been proved \cite{DVB10,LC10} that, when $s_1=s_2=0$,
the zero-discord states have at most one nonzero component of vector $(c_1,c_2,c_3)$,
i.e., all classical-only correlated states lie on the Cartesian axes
$Oc_1$, $Oc_2$ or $Oc_3$.
(This corresponds to the so-called ``Ising spins'' introduced as a matter of fact by
his adviser
W.~Lenz in 1920 \cite{L20,B67}.) 

In the case of Bell-diagonal states, both boundary equations (\ref{eq:SII1})-(\ref{eq:r})
are reduced to a relation
\begin{equation}
   \label{eq:Bd}
   (a-b)^2=(|u|+|v|)^2,
\end{equation}
so that
\begin{equation}
   \label{eq:Bd1}
   2|c_3|=|c_1+c_2|+|c_1-c_2|.
\end{equation}
Thus, the $\pi/2$- and 0-boundaries are coincident, the $Q_\theta$ subdomain
is absent here, and the quantum discord is given by the explicit analytical formula
$Q=\min\{Q_0,Q_{\pi/2}\}$ which is in full agreement with Luo's results \cite{Luo08}.

From Eq.~(\ref{eq:Bd1}), four equations follow
\begin{equation}
   \label{eq:Bd2}
   c_3=\pm c_1\qquad c_3=\pm c_2.
\end{equation}
These planes divide the tetrahedron into subdomains $Q_0$ and $Q_{\pi/2}$,
where the quantum discord takes the values $Q_0$ or $Q_{\pi/2}$.
$Q_0$ subdomain consists of two hexahedrons $(O,v_1,v_2,o_1,o_2)$ and
$(O,v_3,v_4,o_3,o_4)$;
they are shown in Fig.~\ref{fig:6}.
The remaining volume of a tetrahedron belongs to the $Q_{\pi/2}$ states.
It is in two times larger than the volume of $Q_0$ states.

\begin{figure}[t]
\begin{center}
\epsfig{file=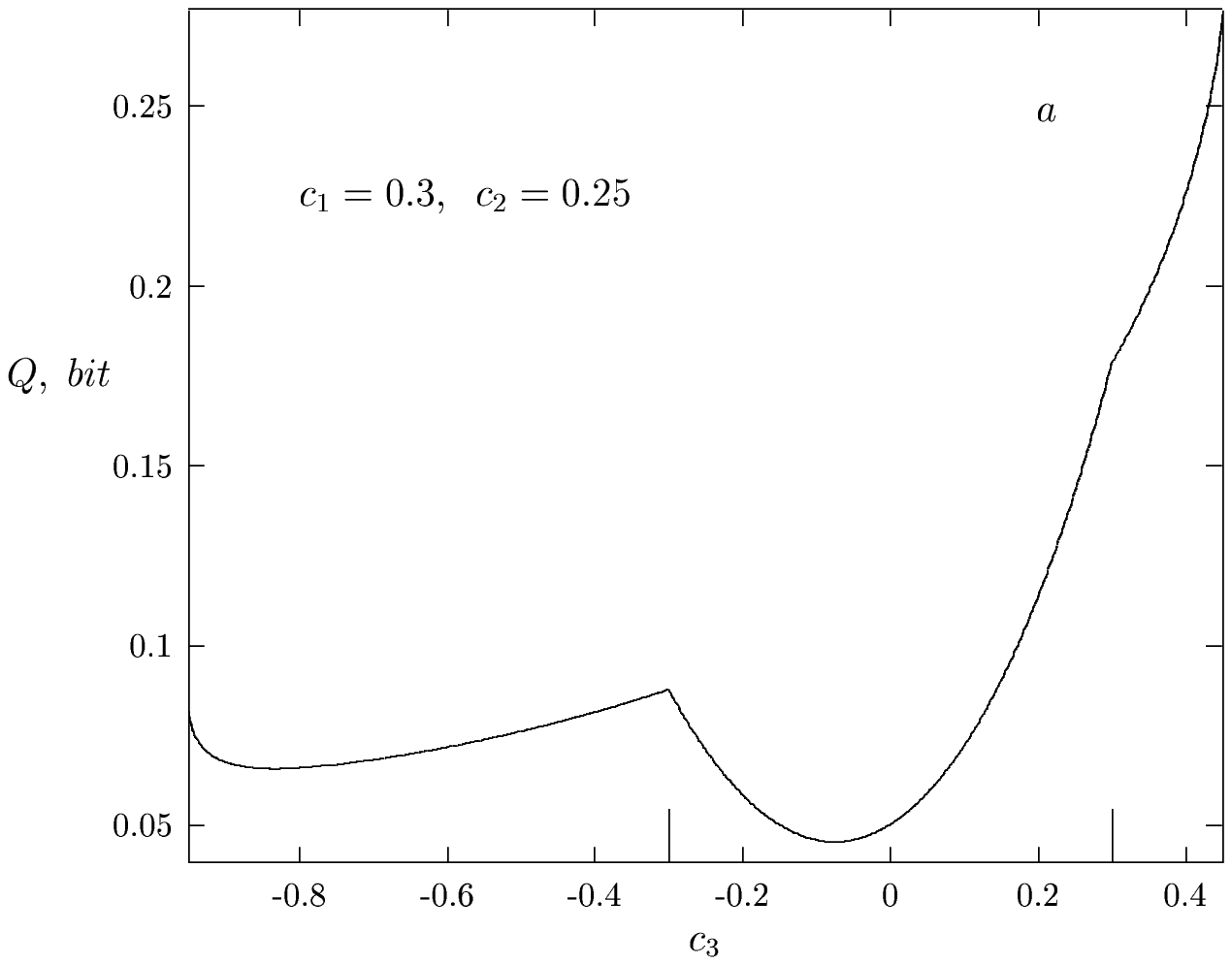,width=5.4cm}
\hspace{0.5cm}
\epsfig{file=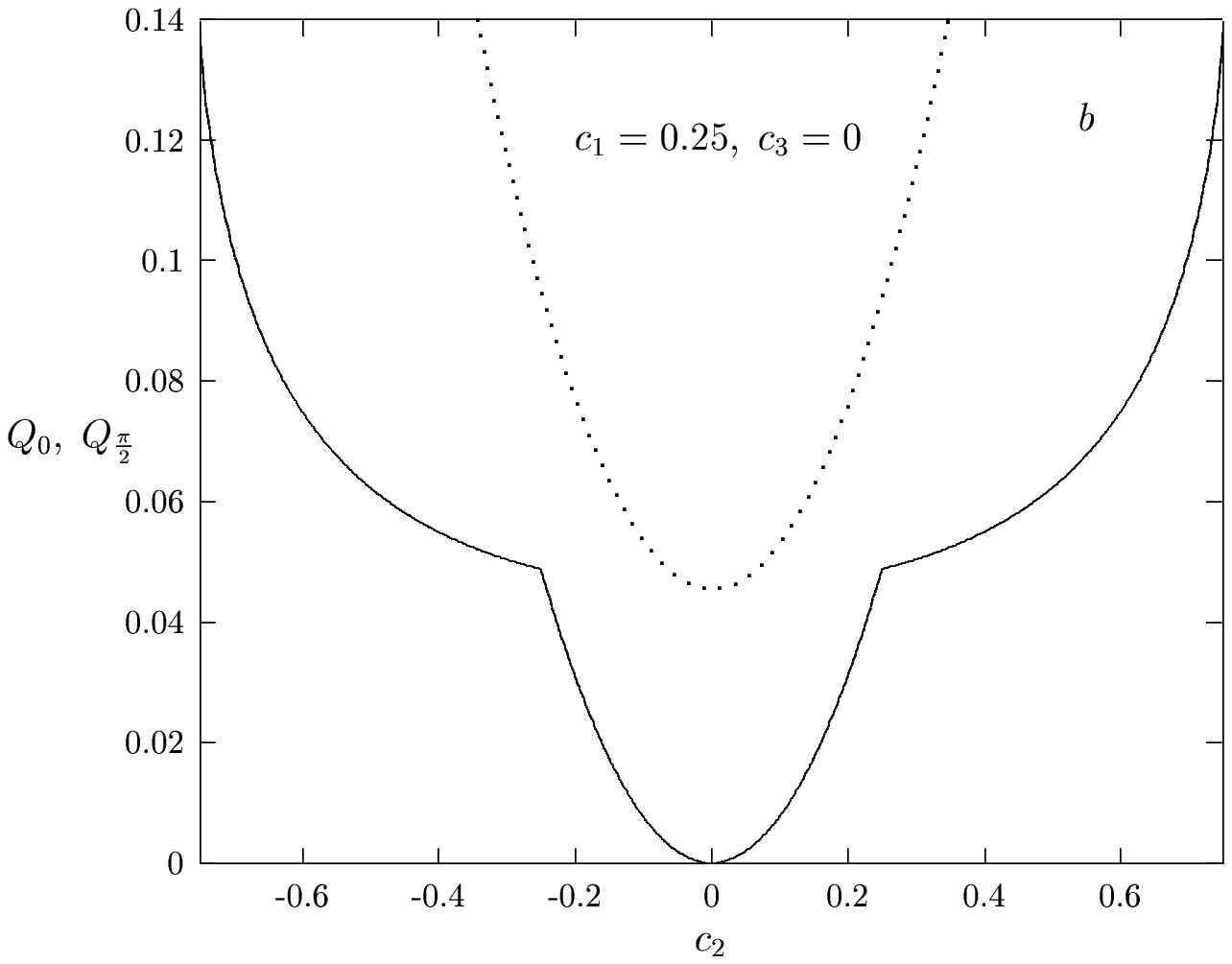,width=5.4cm}
\caption{
Quantum discord for the Bell-diagonal states:
($a$), $Q=\min\{Q_0,Q_{\pi/2}\}$ vs $c_3$ by $c_1=0.3$ and $c_2=0.25$,
longer bars mark the positions of fracture points at $c_3=\pm0.3$;
($b$), $Q=Q_{\pi/2}$ (solid line) and $Q_0$ (dotted line)
vs $c_2$ when $c_1=0.25$ and $c_3=0$
}
\label{fig:7}
\end{center}
\end{figure}

\begin{figure}[b]
\begin{center}
\epsfig{file=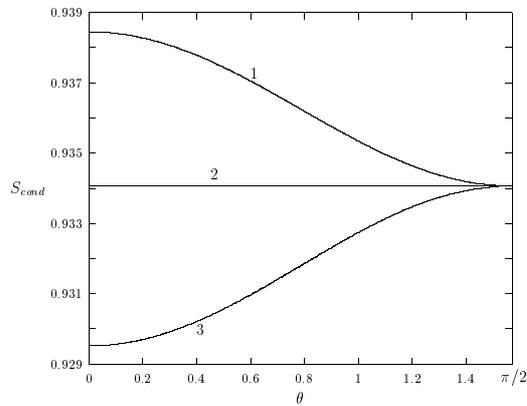,width=6.8cm}
\caption{
Transition between $Q_{\pi/2}$ and $Q_0$ subdomains via a strait line for
the conditional entropy.
Here, $S_{cond}(\theta)$ is at and near the fracture point $c_3=0.3$
on the quantum discord curve in Fig.~\ref{fig:7}~$a$.
The curves 1, 2, and 3 correspond to $c_3=0.29$, 0.3, and 0.31, respectively
}
\label{fig:8}
\end{center}
\end{figure}

The behavior of quantum discord for the Bell-diagonal states along different trajectories
is illustrated in Fig.~\ref{fig:7} by solid lines.
%
Figure~\ref{fig:7}~$a$ shows the discord as a function of $c_3\in[-0.95, 0.45]$
by fixed values of $c_1=0.3$ and $c_2=0.25$.
The curve is continuous but has the fractures at $c_3=\pm0.3$.
They happen when the trajectory crosses the planes dividing
the $Q_0$ and $Q_{\pi/2}$ subdomains (see Fig.~\ref{fig:6}).
In this case, the optimal measurement angle $\theta$ varies discontinuously,
namely, it jumps
from $\theta=0$ to $\theta=\pi/2$ or inversely.
In the vicinity of cross points, the conditional entropy $S_{cond}(\theta)$
changes its form going through a straight line (where any angle
$\theta\in[0,\pi/2]$ is optimal).
Such a regime of conditional entropy behavior is shown in Fig.~\ref{fig:8}.

Figure~\ref{fig:7}~$b$ shows the behavior of branches $Q_0$ and $Q_{\pi/2}$
as functions of $c_2$ by fixed values of other two parameters,
$c_1=0.25$ and $c_3=0$.
Since here $Q_{\pi/2}<Q_0$, the quantum discord $Q$ equals $Q_{\pi/2}$.
The curve $Q_{\pi/2}$ has two fractures.
This means that the branch $Q_{\pi/2}$ is a piecewise-analytic function.
In this case, however, the optimal measurement angle does not change its value
$\theta=\pi/2$
and therefore the position of conditional entropy minimum remains immutable.

\section{Physical systems with the $Q_\theta$ subdomains}
\label{sec:exampl}
We are interested now in the systems with $Q_\theta$ subdomains.
As it was seen from the previous section, such regions do not exist in the
Bell-diagonal states.
Therefore in this section we will consider systems with nonzero Bloch vectors.

\subsection{Phase flip channels}
\label{sec:pfc}
Let us consider the dynamics of quantum discord
under decoherence (for a recent review, see, e.g, \cite{AFA13} and references
therein).
The authors \cite{LWF11} have considered such a dynamics in the phase flip channel.
The problem is to calculate the quantum discord for the $X$ matrix
\begin{eqnarray}
   \label{eq:eps}
   \varepsilon&=&{1\over4}[ 1
   + s_1\sigma_z\otimes1
   + s_21\otimes\sigma_z
   + (1-p)^2c_1\sigma_x\otimes\sigma_x 
   \nonumber\\
   &+& (1-p)^2c_2\sigma_y\otimes\sigma_y 
   + c_3\sigma_z\otimes\sigma_z].
\end{eqnarray}
Here, the parametrized time $p=1-\exp(-\gamma t)$, where $t$ is the time and
$\gamma$ is the phase damping rate.
The authors \cite{LWF11} restricted themselves to the case where
\begin{equation}
   \label{eq:c2s}
   c_2=-c_3c_1,\qquad s_2=c_3s_1,\qquad -1\le c_3\le1,\qquad -1\le s_1\le1 .
\end{equation}
Expansion coefficients in Eq.~(\ref{eq:eps}) are related to the corresponding $X$
matrix elements as
\begin{eqnarray}
   \label{eq:a-v}
   &&a=(1+s_1+s_2+c_3)/4,\quad b=(1+s_1-s_2-c_3)/4,
   \nonumber\\
   &&c=(1-s_1+s_2-c_3)/4,\quad d=(1-s_1-s_2+c_3)/4,
   \\
   &&\ u=(1-p)^2(c_1-c_2)/4,\quad v=(1-p)^2(c_1+c_2)/4 .
   \nonumber
\end{eqnarray}
Owing to the relation $s_2=c_3s_1$, the matrix
elements $a, b, c$, and $d$ satisfy the condition (\ref{eq:acbd}) and hence
the $Q_\theta$ domain is absent here;
conditional entropy behaves similar to that as shown in Fig.~\ref{fig:8}.
Thus, nonzero values of $s_1$ and $s_2$ are the necessary but not sufficient condition
for existence of $Q_\theta$ phase.

\begin{figure}[t]
\begin{center}
\epsfig{file=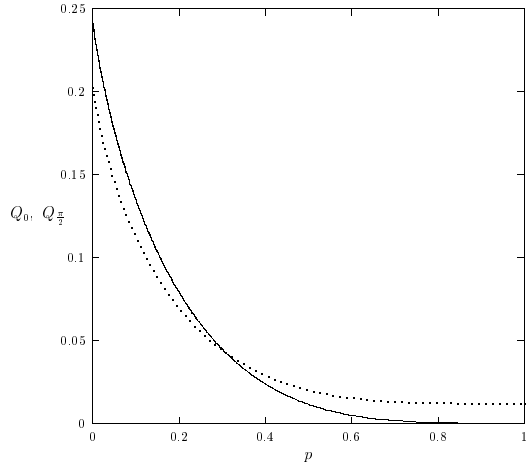,width=6.8cm}
\caption{$Q_0$ (solid line) and $Q_{\pi/2}$ (dotted line) in bits
vs $p$ for the phase flip channel with parameters
$s_1=s_2=0.65$, $c_1=c_2=0.249$, and $c_3=0.5$.
Crossing point of the lines is at $p_\times=0.315\,789\ldots$}
\label{fig:9}
\end{center}
\end{figure}

Consider a different initial state.
For example, let us take
$s_1=s_2=0.65$, $c_1=c_2=0.249$, and $c_3=0.5$.
As seen from Fig.~\ref{fig:9}, the curves $Q_0(p)$ and $Q_{\pi/2}(p)$ have
a crossing point at $p_\times\simeq3158$.
An additional study shows that
the transition $Q_{\pi/2}\rightarrow Q_0$ goes through the appearance of single
minimum on the $S_{cond}(\theta)$ curves inside the interval between $0$ and $\pi/2$
(similarly to the curves on Fig.~\ref{fig:4}).

\begin{figure}[t]
\begin{center}
\epsfig{file=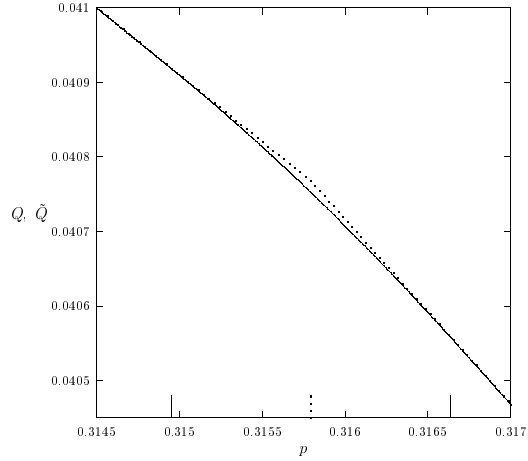,width=6.8cm}
\caption{Dependencies of the false discord $\tilde Q=\min\{Q_{\pi/2},Q_0\}$
(dotted line) and
the corrected quantum discord $Q=\min\{Q_{\pi/2},Q_\theta,Q_0\}$ (solid line)
vs $p$ for the phase flip channel with parameters $s_1=s_2=0.65$,
$c_1=c_2=0.249$, and $c_3=0.5$.
Longer solid bars mark the boundaries $p_{\pi/2}=0.314\,949$ and $p_0=0.316\,637$.
Longer dotted bar marks the position of a fracture, $p_\times=0.315\,789$,
on the  curve $\tilde Q(p)$}
\label{fig:10}
\end{center}
\end{figure}

Solution of equations for the boundaries, Eqs.~(\ref{eq:SII1})-(\ref{eq:r}),
shows that the $\pi/2$- and 0-boundaries do not coincide now and therefore
the $Q_\theta$ region exists here (see  Fig.~\ref{fig:10}).

\subsection{Thermal discord}
\label{sec:XYZ}

We now discuss systems at thermal equilibrium.
Let us consider the XYZ spin Hamiltonian
\begin{equation}
   \label{eq:Hxyz}
   {\cal H} = -\frac{1}{2}(J_x\sigma_1^x\sigma_2^x 
   + J_y\sigma_1^y\sigma_2^y 
   + J_z\sigma_1^z\sigma_2^z
	 + B_1\sigma_1^z + B_2\sigma_2^z).
\end{equation}
This Hamiltonian contains five independent parameters
$J_x,J_y,J_z,B_1,B_2\in(-\infty,\infty)$ (i.e., in ${\cal R}^5$) and
is the most general real symmetric traceless $X$ matrix.
The corresponding Gibbs density matrix is given as
\begin{equation}
   \label{eq:rho-G}
   \rho_{AB}=\frac{1}{Z}e^{-\beta{\cal H}}
\end{equation}
(here $\beta=1/T$, $T$ is the temperature in energy units, $Z$
is the partition function) and has also the five-parameter real $X$ structure.
Thus, the map
$( B_1/T, B_2/T, J_x/T, J_y/T, J_z/T)\leftrightarrow(s_1, s_2, c_1, c_2, c_3)$
(that is ${\cal R}^5\leftrightarrow{\cal D}$)
allows in general to change the density-matrix language on
a picture of interactions in the XYZ dimer in inhomogeneous
fields $B_1$ and $B_2$. 

Having solved eigenproblem for the Hamiltonian (\ref{eq:Hxyz}) we then find expressions
for the thermal density matrix elements
\begin{eqnarray}
   \label{eq:avT}
   &&a={1\over2}\frac{\cosh(\beta R_1/2)+[(B_1+B_2)/R_1]\sinh(\beta R_1/2)}
	                 {\cosh(\beta R_1/2)+\exp(-\beta J_z)\cosh(\beta R_2/2)},
   \nonumber\\
   &&b={1\over2}\frac{\cosh(\beta R_2/2)+[(B_1-B_2)/R_2]\sinh(\beta R_2/2)}
	                 {\exp(\beta J_z)\cosh(\beta R_1/2)+\cosh(\beta R_2/2)},
   \nonumber\\
   &&c={1\over2}\frac{\cosh(\beta R_2/2)-[(B_1-B_2)/R_2]\sinh(\beta R_2/2)}
	                 {\exp(\beta J_z)\cosh(\beta R_1/2)+\cosh(\beta R_2/2)},\\
   &&d={1\over2}\frac{\cosh(\beta R_1/2)-[(B_1+B_2)/R_1]\sinh(\beta R_1/2)}
	                 {\cosh(\beta R_1/2)+\exp(-\beta J_z)\cosh(\beta R_2/2)},
   \nonumber\\
   &&u={1\over2}\frac{[(J_x-J_y)/R_1]\sinh(\beta R_1/2)}
	                 {\cosh(\beta R_1/2)+\exp(-\beta J_z)\cosh(\beta R_2/2)},
   \nonumber\\
   &&v={1\over2}\frac{[(J_x+J_y)/R_2]\sinh(\beta R_2/2)}
	                 {\exp(\beta J_z)\cosh(\beta R_1/2)+\cosh(\beta R_2/2)},
   \nonumber
\end{eqnarray}
where
\begin{equation}
   \label{eq:R1R2}
   R_1=[(B_1+B_2)^2+(J_x-J_y)^2]^{1/2},\quad
   R_2=[(B_1-B_2)^2+(J_x+J_y)^2]^{1/2}.
\end{equation}
For the correlations functions (\ref{eq:s1-c3}), we have respectively
\begin{eqnarray}
   \label{eq:csT}
	 &&c_{1,2}={2\over Z}\lbrack\!\lbrack\pm[(J_x-J_y)/R_1]e^{\beta J_z/2}\sinh(\beta R_1/2)
   \nonumber\\
   &&\qquad\qquad+[(J_x+J_y)/R_2]e^{-\beta J_z/2}\sinh(\beta R_2/2)\rbrack\!\rbrack,
   \nonumber\\
   &&c_3={2\over Z}[e^{\beta J_z/2}\cosh(\beta R_1/2)-e^{-\beta J_z/2}\cosh(\beta R_2/2)],\\
	 &&s_{1,2}={2\over Z}\lbrack\!\lbrack[(B_1+B_2)/R_1]e^{\beta J_z/2}\sinh(\beta R_1/2)
   \nonumber\\
	 &&\qquad\qquad\pm[(B_1-B_2)/R_2]e^{-\beta J_z/2}\sinh(\beta R_2/2)\rbrack\!\rbrack,
   \nonumber
\end{eqnarray}
where the partition function equals
\begin{equation}
   \label{eq:ZT}
   Z=2[e^{\beta J_z/2}\cosh(\beta R_1/2)+e^{-\beta J_z/2}\cosh(\beta R_2/2)]
\end{equation}
and $R_1$ and $R_2$ are given again by Eq.~(\ref{eq:R1R2}).

For each choice of interaction constants $J_x$, $J_y$, $J_z$ and external fields
$B_1$ and $B_2$ we will find the points where the condition $Q_0=Q_{\pi/2}$
is satisfied.
After this we will again study the changes of curves $S_{cond}(\theta)$ in the
neighborhood of points found.

\begin{figure}[t]
\begin{center}
\epsfig{file=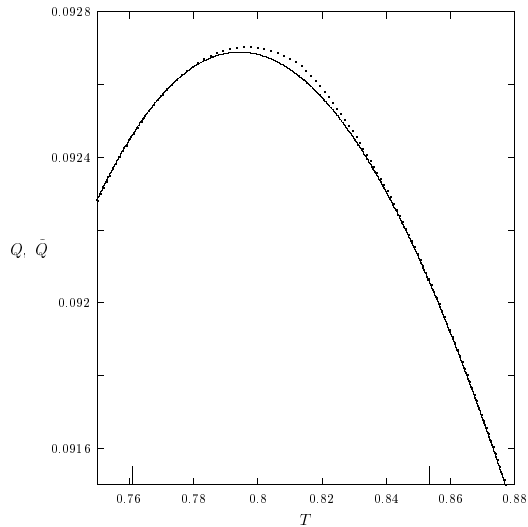,width=6.8cm}
\caption{Dependencies of the false discord $\tilde Q=\min\{Q_{\pi/2},Q_0\}$
(dotted line) and
the correct quantum discord $Q=\min\{Q_{\pi/2},Q_\theta,Q_0\}$ (solid line)
for the XXZ dimer with parameters $J=1$, $J_z=1.02$ and $B=1$.
Longer bars mark the temperatures $T_{\pi/2}=0.76106$ and $T_0=0.85361$.
Domains $T\le T_{\pi/2}$, $T_{\pi/2}<T<T_0$, and $T\ge T_0$ correspond
to the discord branches $Q_{\pi/2}$, $Q_\theta$, and $Q_0$, respectively}
\label{fig:11}
\end{center}
\end{figure}

Taking, for example, a dimer with parameters $J_x=J_y=J=1$, $J_z=1.02$,
and $B_1=B_2=B=1$ (that is the XXZ dimer in an uniform field) we consider
the thermal discord behavior by a transition
from the subdomain $Q_{\pi/2}$ to $Q_0$ one (Fig.~\ref{fig:11}).
From the figure one can see that down to the crossing point $T_\times=0.81296$ the
discord ${\tilde Q}$, according to Refs.~\cite{ARA10,FWBAC10,LWF11}, equals
$Q_{\pi/2}$ and above the point $T_\times$ it equals $Q_0$.
If this were valid, the discord ${\tilde Q}=\min\{Q_0,Q_{\pi/2}\}$
would have a fracture at the intersection point $T_\times$.
However, in fact, the true discord $Q$ is a smooth function
(at least, it is a function of differentiability class $C^1$).
This follows from the numerical solution of the task in the intermediate domain,
where the $S_{cond}(\theta)$ curves change similar as in Fig.~\ref{fig:4}.
Results for the quantum discord are shown again in Fig.~\ref{fig:11}.
At the bifurcations points $T_{\pi/2}=0.76106$ and $T_0=0.85361$,
the higher derivatives of quantum discord $Q=\min\{Q_{\pi/2},Q_\theta,Q_0\}$ exhibit
a discontinuous behavior.

\subsection{Heteronuclear systems with dipolar coupling}
\label{sec:hetero}
Let us consider the system (\ref{eq:Hxyz}) with parameters $J_x=J_y=-D$
and $J_z=2D$.
Such a model corresponds to a dipolar coupled dimer which is stretched along
the $z$ axis \cite{KY13}
\begin{equation}
   \label{eq:Hdd}
   {\cal H} = {1\over2}D(\sigma_1^x\sigma_2^x 
   + \sigma_1^y\sigma_2^y 
   - 2\sigma_1^z\sigma_2^z)
	 - {1\over2}(B_1\sigma_1^z + B_2\sigma_2^z).
\end{equation}
Here the dipolar coupling constant (in frequency units) equals
\begin{equation}
   \label{eq:D}
   D = \frac{\mu_0}{4\pi}\frac{\gamma_1\gamma_2}{2r_0^3}, 
\end{equation}
where $\mu_0$ is the magnetic permeability of free space,
$\gamma_1$ and $\gamma_2$ are the gyromagnetic ratios of particles in the dimer,
and $r_0$ is the distance between those particles.
Normalized fields $B_1$ and $B_2$ in Eq.~(\ref{eq:D}) are
\begin{equation}
   \label{eq:B1B2}
   B_1 = \gamma_1 B_0, \qquad B_2=\gamma_2 B_0, 
\end{equation}
where $B_0$ is the external magnetic field induction.

\begin{figure}[t]
\begin{center}
\epsfig{file=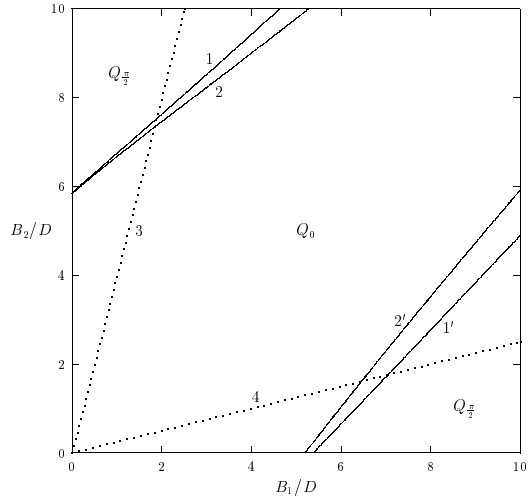,width=6.8cm}
\caption{
Subdomains $Q_{\pi/2}$, $Q_0$, and (between the lines $1,2$
and $1^\prime,2^\prime$) $Q_\theta$ for the spin dimer (\ref{eq:Hdd})
at the normalized temperature $T/D=1$.
Dotted lines 3 and 4 correspond to $B_2=4B_1$ and $B_2=B_1/4$, respectively}
\label{fig:12}
\end{center}
\end{figure}

We have performed necessary calculations (according to our approach developed in the
previous sections) and found the subdomains of quantum discord in the plane
$(B_1/D,B_2/D)$.
The results are shown at the normalized temperature $T/D=1$ in Fig.~\ref{fig:12}.
From this figure one can see that such a system has the $Q_\theta$ regions
(between the $1,2$ and $1^\prime,2^\prime$ lines).
These regions can be reached by varying the external magnetic field $ B_0$.
Two possible trajectories are shown in Fig.~\ref{fig:12} by dotted lines,
$B_2=4B_1$ and $B_2=B_1/4$.
(The value $\gamma_2/\gamma_1=4$ approximately corresponds to
the quotient of gyromagnetic ratios for the nucleus of $^1\!$H and $^{13}$C.)

We found also that in the $Q_\theta$ subdomain the conditional entropy
$S_{cond}(\theta)$ has only one minimum that is located in the interval
$(0,\pi/2)$.
The picture is qualitatively similar to that is shown in Fig.~\ref{fig:4}.

So, the $Q_\theta$ region and corresponding sudden changes of quantum
correlation behavior at their boundaries can be observed in solid materials
with nuclear dimers.

\section{Results and perspectives}
\label{sec:concl}
In light of the above, the calculation of quantum discord of any $X$
states can be achieved by following steps.
First, the density matrix (\ref{eq:rho-Xc}) is transformed to
the real form, i.e., the quantities $u$ and $v$ are calculated
using Eqs.~(\ref{eq:u}) and (\ref{eq:v}).
It is also well to solve the equation $Q_0=Q_{\pi/2}$ and determine
possible crossing points of branches $Q_0$ and $Q_{\pi/2}$.
Then the equations $S_{cond}^{\prime\prime}(0)=0$ and
$S_{cond}^{\prime\prime}(\pi/2)=0$ are solved to find the boundaries
for the intermediate subdomain $Q_{\theta}$.
After this, one should numerically find the optimal
measurement angle $\theta\in(0,\pi/2)$ and compute $Q_{\theta}=Q(\theta)$.
As a result, the quantum discord is given by $Q=\min\{Q_0, Q_\theta, Q_{\pi/2}\}$.

The formula for calculating the quantum discord belongs to a
piecewise-defined type
\begin{equation}
   \label{eq:f}
   f(x)=
	 \cases{
      F(x,a), &$x\in\Omega_a$\cr
	    F(x,b), &$x\in\Omega_b$\cr
	    \min_{\alpha\in(a,b)} F(x,\alpha), &$x\in\Omega_c$.
		}
\end{equation}
In other words, the domain of definition, $\Omega$, of the function $f(x)$
consists of subdomains in which the function is given by closed
analytical expressions or it exists in a numerical form.

So, the quantum discord of $X$ states is represented analytically
if the $Q_\theta$ subdomain is absent.
Then the quantum discord is given by the closed form $Q=\min\{Q_{\pi/2},Q_0\}$.
The discord is continuous, but generally speaking it is a piecewise smooth function.
In particular, this is valid for a spacial class of $X$ states, namely,
for the Bell-diagonal states.
For them, we found the $Q_0$ and $Q_{\pi/2}$ regions in the total domain of their
definition (Fig.~\ref{fig:6}).
It would be interesting to find the subdomains $Q_0$, $Q_\theta$, and $Q_{\pi/2}$
in the five-dimensional domain ${\cal D}$ making, e.g., an atlas of maps.

Also, we have shown in this paper that the boundaries for the transition
subdomain from $Q_0$ to $Q_{\pi/2}$ or reversely are exactly defined.
They consist of nonanalyticity points which are bifurcation ones.
The corresponding equations for these boundaries have been found.
The boundaries may coincide and then the quantum
discord is evaluated analytically in the total domain of definition.
The regions $Q_\theta$ with the optimal intermediate angles $\theta\in(0,\pi/2)$
have been found for a number of physical systems including the phase flip channels,
spin dimers at the thermal equilibrium, heteronuclear systems with dipolar interaction.
The transitions
$Q_{\pi/2}\leftrightarrow Q_\theta\leftrightarrow Q_0$
occur continuously and smoothly.
This is a new type of transitions for the quantum discord.

We have found only two regimes for the conditional entropy change
by above transitions:
(i) via the birth of one intermediate minimum (as shown in
Fig.~\ref{fig:4}) and
(ii) via the strait line (as shown in Fig.~\ref{fig:8}).
It is hoped that our observations will be rigorously proofed and, maybe,
generalized in the future.

At present the attempts are made to obtain analytical formulas for the super quantum
discord of $X$ states with nonzero Bloch vectors \cite{EF14,LMWFW15}.
In this connection one should note the the authors do not take into
account a possibility of intermediate optimal angles for the weak measurements
which are a generalization of projective ones.


\section*{Acknowledgements}
\label{sec:ackn}
The author thanks A.~I.~Zenchuk for valuable remarks.
The research was supported by the RFBR grants Nos.~13-03-00017 and 15-07-07928
and by the program No.~8 of the Presidium of RAS.




\begin{thebibliography}{99}

\bibitem{AFOV08}
Amico,~L., Fazio,~R., Osterloh,~A., Vedral,~V.:
Entanglement in many-body systems.
Rev. Mod. Phys. {\bf 80}, 517 (2008)

\bibitem{HHHH09}
Horodecki,~R., Horodecki,~P., Horodecki,~M., Horodecki,~K.:
Quntum entanglement.
Rev. Mod. Phys. {\bf 81}, 865 (2009)

\bibitem{CMS11}
C$\acute{\rm e}$leri,~L.~C., Maziero,~J., Serra,~R.~M.:
Theoretical and experimental aspects of quantum discord and related measures.
Int. J. Quant. Inf. {\bf 11}, 1837 (2011)

\bibitem{MBCPV12}
Modi,~K., Brodutch,~A., Cable,~H., Paterek,~T., Vedral,~V.:
The classical-quantum boundary for correlations: discord and related measures.
Rev. Mod. Phys. {\bf 84}, 1655 (2012)

\bibitem{AFY14}
Aldoshin,~S.~M., Fel'dman,~E.~B., Yurishchev,~M.~A.:
Quantum entanglement and quantum discord in magnetoactive materials (Review Article).
Fiz. Nizk. Temp. {\bf 40}, 5 (2014) (in Russian); 
Low Temp. Phys. {\bf 40}, 3 (2014)

\bibitem{H14}
Huang,~Y.:
Computing quantum discord is NP-coplete.
New J. Phys. {\bf 16}, 033027 (2014)

\bibitem{W98}
Hill,~S., Wootters,~W.~K.:
Entanglement of a pair of quantum bits.
Phys. Rev. Lett. {\bf 78}, 5022 (1997);
Wootters,~W.~K.:
Entanglement of formation of an arbitrary state of two qubits.
Phys. Rev. Lett. {\bf 80}, 2245 (1998);
Verstraete,~F., Dehaene,~J., De~Moor,~B.:
Local filtering operations on two qubits.
Phys. Rev. A {\bf 64}, 010101(R) (2001);
Audenaert,~K., Verstraete,~F., De~Moor,~B.:
Variational characterizations of separability and entanglement of formation.
Phys. Rev. A {\bf 64}, 052304 (2001)

\bibitem{Luo08}
Luo,~S.:
Quantum discord for two-qubit systems.
Phys. Rev. A {\bf 77}, 042303 (2008)

\bibitem{ARA10}
Ali,~M., Rau,~A.~R.~P., Alber,~G.:
Quantum discord for two-qubit $X$ states.
Phys. Rev. A {\bf 81}, 042105 (2010);
Erratum in: Phys. Rev. A {\bf 82}, 069902(E) (2010)

\bibitem{FWBAC10}
Fanchini,~F.~F., Werlang,~T., Brasil,~C.~A., Arruda,~L.~G.~E., Caldeira,~A.~O.:
Non-Markovian dynamics of quantum discord.
Phys. Rev. A {\bf 81}, 052107 (2010)

\bibitem{LWF11}
Li,~B., Wang,~Z.-X., Fei,~S.-M.:
Quantum discord and geometry for a class of two-qubit states.
Phys. Rev. A {\bf 83}, 022321 (2011)

\bibitem{DWZ11}
Ding,~B.-F., Wang,~X.-Y., Zhao,~H.-P.:
Quantum and classical correlations for a two-qubit $X$ structure density matrix.
Chin. Phys. B {\bf 20}, 100302 (2011)

\bibitem{VR12}
Vinjanampathy,~S., Rau,~A.~R.~P.:
Quantum discord for qubit-qudit systems.
J. Phys. A: Math. Theor. {\bf 45}, 095303 (2012)

\bibitem{YE07}
Yu,~T., Eberly,~T.~H.:
Evolution from entanglement to decoherence of bipartite mixed ''X`` states.
Quant. Inf. Comput {\bf 7}, 459 (2007)

\bibitem{R09}
Rau,~A.~R.~P.:
Algebraic characterization of $X$-states in quantum information.
J. Phys. A: Math. Theor. {\bf 42}, 412002 (2009)

\bibitem{LMXW11}
Lu,~X.-M., Ma,~J., Xi,~Z., Wang,~X.:
Optimal measurements to access classical correlations of two-qubit states.
Phys. Rev. A {\bf 83}, 012327 (2011)

\bibitem{CZYYO11}
Chen,~Q., Zhang,~C., Yu,~S., Yi,~X.~X., Oh,~C.~H.:
Quantum discord of two-qubit $X$ states.
Phys. Rev. A {\bf 84}, 042313 (2011)

\bibitem{H13}
Huang,~Y.:
Quantum discord for two-qubit $X$ states: analytical formula with very small worst-case error.
Phys. Rev. A {\bf 88}, 014302 (2013)

\bibitem{CRC10}
Ciliberti,~L., Rossignoli,~R., Canosa,~N.:
Quantum discord in finite $XY$ chains.
Phys. Rev. A {\bf 82}, 042316 (2010)

\bibitem{Y14}
Yurischev,~M.~A.:
Quantum discord for general X and CS states: a piecewise-analytic-numerical formula.
arXiv:1404.5735v1 [quant-ph]

\bibitem{Y14a}
Yurishchev,~M.~A.:
NMR dynamics of quantum discord for spin-carrying gas molecules in a closed nanopore.
J. Exp. Theor. Phys. {\bf 119}, 828 (2014)

\bibitem{KHJP10}
Kim~H., Hwang~M.-R., Jung~E., Park~D.K.:
Difficulties in analytic computation for relative entropy of entanglement.
Phys. Rev. A {\bf 81}, 052325 (2010)

\bibitem{OZ01}
Ollivier,~H., ~Zurek,~W.~H.:
Quantum discord: a measure of the quantumness of correlations.
Phys. Rev. Lett. {\bf 88}, 017901 (2001)

\bibitem{Z03}
Zurek,~W.~H.:
Quantum discord and Maxwell's demons.
Phys. Rev. A {\bf 67}, 012320 (2003)


\bibitem{PKF13}
Pinto,~J.~P.~G., Karpat,~G., Fanchini,~F.~F.:
Sudden change of quantum discord for a system of two qubits.
Phys. Rev. A {\bf 88}, 034304 (2013)

\bibitem{GGZ11}
Galve,~F., Giorgi,~G.~L., Zambrini,~R.:
Maximally discordant mixed states of two qubits.
Phys. Rev. A {\bf 83}, 012102 (2011)

\bibitem{A04}
Arnold,~V.~I.:
Catastrophe theory.
Springer-Verlag, Berlin, Heidelberg, New York (1992), sec.~10

\bibitem{HH96}
Horodecki~R., Horodecki~M.:
Information-theoretic aspects of inseparability of mixed states.
Phys. Rev. A {\bf 54}, 1838 (1996)

\bibitem{FACCA10}
Ferraro~A., Aolita~L., Cavalcanti~D., Cucchietti~F.~M., Ac${\acute{\rm i}}$n~A.:
Almost all quantum states have nonclassical correlations.
Phys. Rev. A {\bf 81}, 052318 (2010)

\bibitem{DVB10}
Daki${\acute {\rm c}}$~B., Vedral~V., Brucner~${\breve{\rm C}}$.:
Necessary and sufficient condition for nonzero quantum discord.
Phys. Rev. Lett. {\bf 105}, 190502 (2010)

\bibitem{LC10}
Lang~M.~D., Caves~C.~M.:
Quantum discord and the geometry of Bell-diagonal states.
Phys. Rev. Lett. {\bf 105}, 150501 (2010)

\bibitem{L20}
Lenz~W.:
Beitrag zum Verst${\ddot{\rm a}}$ndnis der magnetischen Erscheinungen in festen
K${\ddot{\rm o}}$rpern.
Phys. Z. {\bf 21}, 613 (1920)

\bibitem{B67}
Brush~S.~G.:
History of the Lenz--Ising model.
Rev. Mod. Phys. {\bf 39}, 883 (1967)

\bibitem{AFA13}
Aaronson,~B., Lo~Franco,~R., Adesso,~G.:
Comparative investigation of the freezing phenomena for quantum
correlations under nondissipative decoherence.
Phys. Rev. A {\bf 88}, 012120 (2013)


\bibitem{KY13}
Kuznetsova,~E.~I., Yurischev,~M.~A.:
Quantum discord in spin systems with dipole-dipole interaction.
Quantum Inf. Process. {\bf 12}, 3587 (2013)

\bibitem{EF14}
Eftekhari,~H., Faizi,~E.:
Super quantum discord for a class of two-qubit states with weak measurement.
arXiv:1409.4329v1 [quant-ph]

\bibitem{LMWFW15}
Li,~T., Ma,~T., Wang,~Y., Fei,~S., Wang.~Z.:
Super quantum discord for X-type states.
Int. J. Theor. Phys. {\bf 54}, 680 (2015)

\end{thebibliography}
\end{document}